\documentclass{article}

 \usepackage[main, final]{neurips_2026}
 \usepackage{flafter}
 \usepackage{placeins}

\usepackage[utf8]{inputenc} 
\usepackage[T1]{fontenc}    
\usepackage{hyperref}       
\usepackage{url}            
\usepackage{booktabs}       
\usepackage{amsfonts}       
\usepackage{nicefrac}       
\usepackage{microtype}      
\usepackage{xcolor}         
\usepackage{graphicx}
\usepackage{amsmath}

\title{Pushing the Frontier of Full-Song Generation: Hierarchical Autoregressive Planning Meets Flow-Matching Rendering}

\author{%
      \parbox{0.96\textwidth}{%
        \centering
        {\small\normalfont
        \textbf{\textit{(Equal contribution; alphabetical by family name.)}}\\[-0.1em]
        Junyu Dai, Xinyue Fan, Weiqin Li, Xiangang Li, Yunjia Li, Bin Ma,\\
        Yukun Ma, Chongjia Ni, Yufei Shi, Biao Tian, Haoxu Wang, Menglin Wu, Jianwei Yu,\\
        Huaicheng Zhang, Han Zhao, Shengkui Zhao, and Haina Zhu
        }\\[0.55em]
        {\small\normalfont
        \raisebox{-0.15\height}{\includegraphics[height=1.05em]{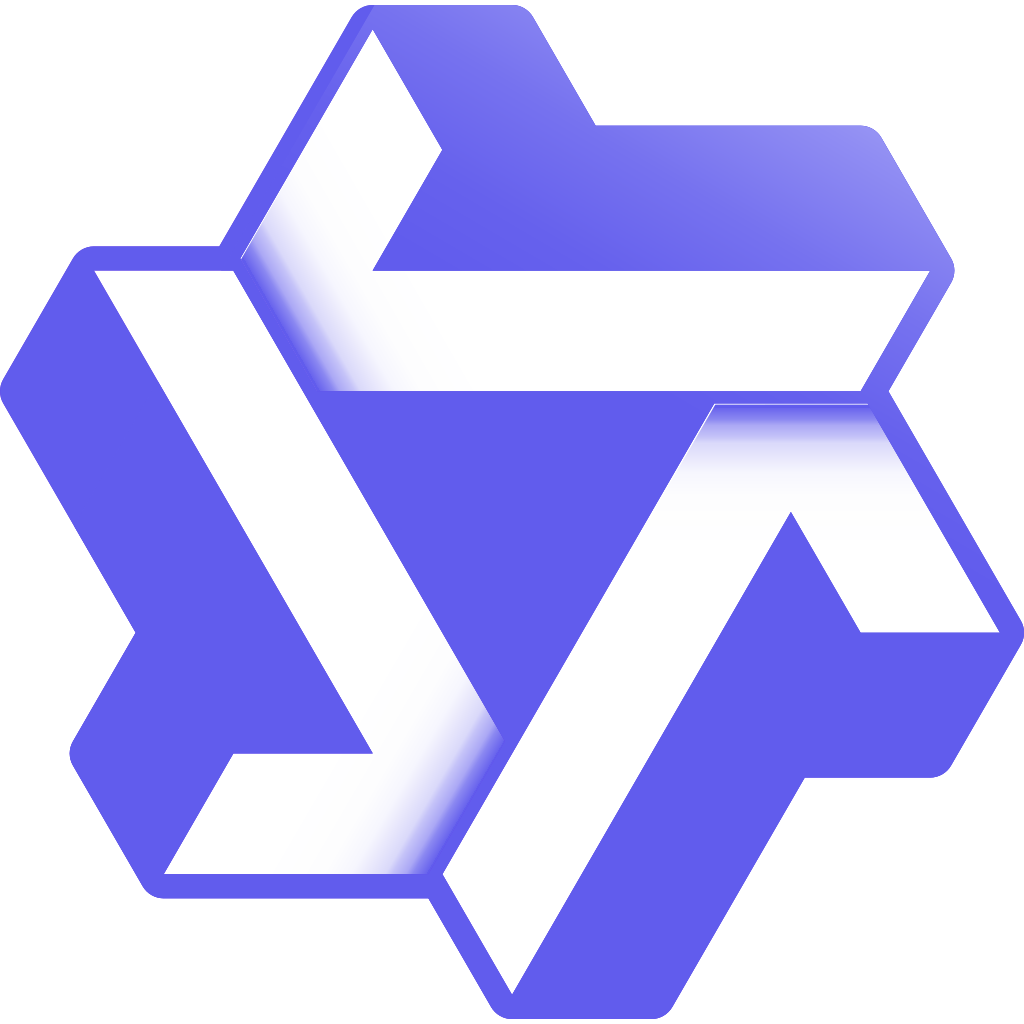}}%
        \hspace{0.40em}\textbf{Alibaba Token Foundry}}\\[0.55em]
        {\small\normalfont
        \href{https://hifi-song-generation.github.io/}
        {\textcolor{blue}{\underline{Demo Page}}}}
      }
  }

\begin{document}


\maketitle

\begin{abstract}
  In this report, we present a unified song generation framework capable of producing high-quality full-length music from lyrics, text descriptions, and musical attributes. The proposed framework supports three tasks: Lyrics-to-Song Generation, which generates complete songs from text descriptions, lyrics, and musical attributes; Instrumental Music Generation, which creates music without vocals; and Cover Song Generation, which reinterprets existing songs with different styles while preserving their melodic content. Architecturally, our system consists of four main components: a semantic-aware tokenizer, hybrid-LM, FullDiT, and a two-level melody module. The tokenizer encodes audio into 8-codebook RVQ tokens for efficient discrete music representation. Based on these tokens, hybrid-LM performs hierarchical autoregressive audio-token modeling for full-song generation. To improve audio fidelity, FullDiT performs full-song flow matching in a continuous VAE latent space conditioned on codec tokens, lyrics, and text captions. For cover song generation, the melody module extracts and discretizes melody cues from reference audio to guide generation while preserving the original melodic content. Finally, we investigate DPO, GRPO, and OPD as reward-based post-training strategies for hybrid-LM and apply flow-based GRPO to FullDiT to improve musicality and rendering quality. Experimental results on a multilingual automatic benchmark, complemented by the Artificial Analysis Music with Vocals leaderboard, show that the proposed framework achieves competitive performance in the evaluated settings.
\end{abstract}

\section{Introduction}

Music generation has recently emerged as a central challenge in generative modeling, aiming to synthesize music that is structurally coherent, acoustically natural, and consistent with desired musical attributes \cite{lei2026levo, lei2026levo2, gong2025ace, gong2026ace, ning2025diffrhythm, yang2026heartmula, yang2026songbloom, yuan2025yue}. Compared with general audio generation, full-song generation is considerably more demanding, as music comprises multiple interacting elements, including lyrics, vocal melody, accompaniment, instrumental arrangements, rhythm, and long-range structural organization.  A practical system should support not only open-ended song generation from lyrics and text descriptions, but also instrumental music generation \cite{copet2023simple, chen2024musicldm, agostinelli2023musiclm} and cover song generation \cite{li2026songecho} with controllable style and melodic characteristics. Despite rapid progress in music generation, producing full-length songs with clear vocals, natural accompaniment, and high perceptual quality remains difficult.

\begin{figure}[!t]
    \centering
    \includegraphics[width=0.9\linewidth]{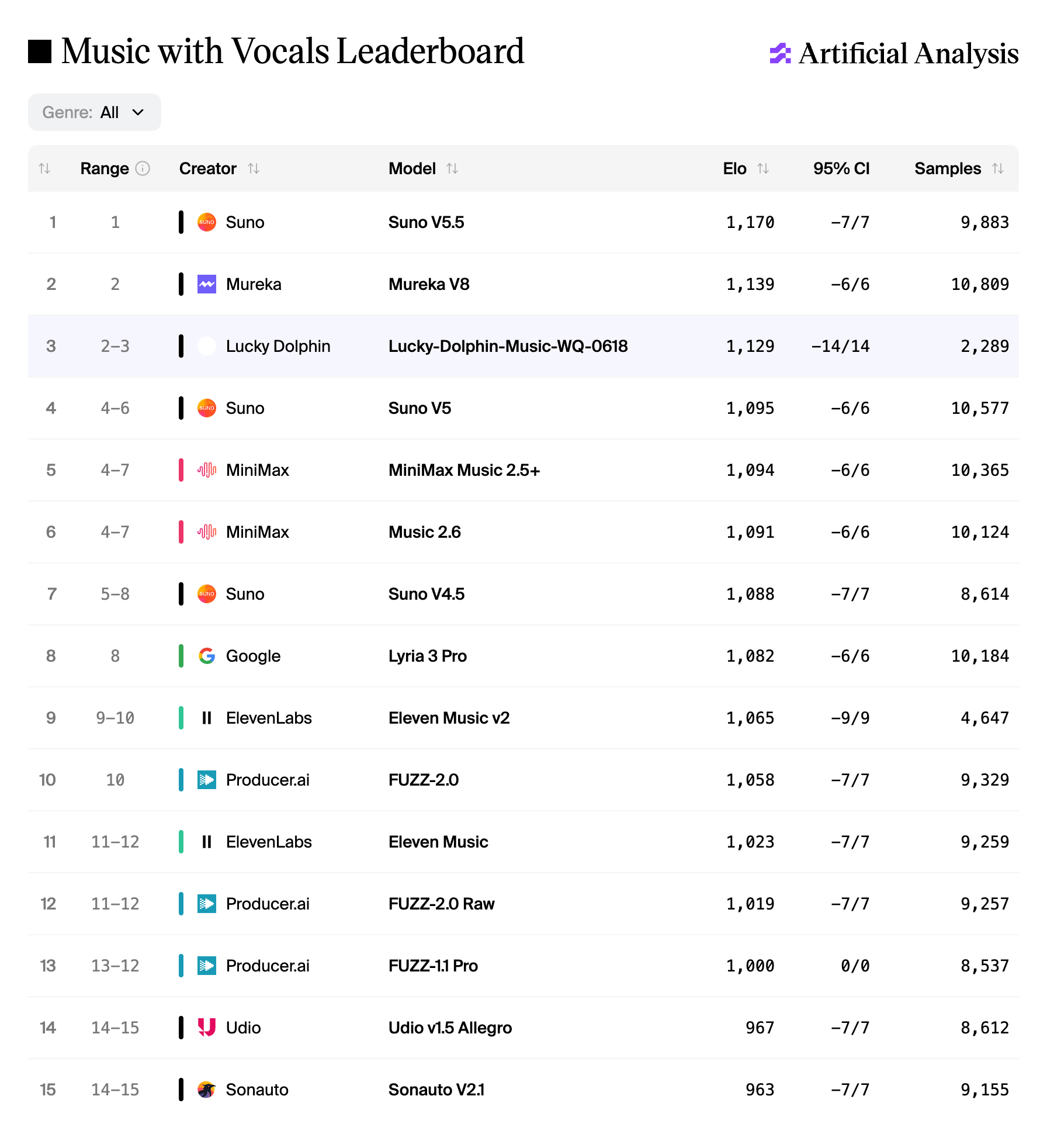}
    \caption{Results on the Artificial Analysis Music with Vocals leaderboard. The leaderboard entry \emph{Lucky Dolphin} corresponds to the proposed system, which was submitted anonymously for evaluation. It achieves an Elo rating of 1,129 and an official rank range of 2--3. Error bars denote 95\% confidence intervals.}
    \label{fig:aa_leaderboard}
\end{figure}

One core difficulty of music generation lies in its highly compositional nature. A full song often contains multiple instruments, overlapping melodies, and rich temporal variation, making it difficult to model music effectively with overly compressed discrete representations. In particular, single-codebook tokenization is often inadequate for representing the diversity and complexity of musical content, which has led many approaches to model vocals and accompaniment separately \cite{lin2025duo, yuan2025yue, lei2026levo, lei2026levo2, xu2025mucodec}. By comparison, RVQ-based multi-codebook modeling is better suited to capturing fine-grained audio details \cite{liu2026khala, yang2026heartmula}. However, using multiple codebooks also makes prediction harder, as higher codebooks are more difficult to predict accurately. 

At the same time, music generation places a high demand on perceptual quality: beyond correct composition, the system must produce realistic singing voices, natural accompaniment, and high-fidelity audio over full-song duration. Existing chunk-wise decoding approaches \cite{lei2026levo, yang2026songbloom} for diffusion or DiT-based synthesis are effective for local generation, but can limit global consistency and ultimate audio quality when applied to long-form music. 

In addition, cover song generation further requires the model to preserve the main melody of a reference song in the generated output. Although prior studies and tools \cite{li2026songecho, wei2023rmvpe, chang2024yourmt3} have developed various approaches for vocal melody extraction, balancing the note-level melodic structure that defines the song with the fine-grained pitch variations that characterize natural singing remains an open challenge.

In this report, we present a unified framework for controllable full-song generation. The proposed framework supports three core tasks: Lyrics-to-Song Generation, Instrumental Music Generation, and Cover Song Generation. The overall system is built upon four main components: a semantic-aware tokenizer, hybrid-LM, a two-level melody module, and FullDiT. Given lyrics, text descriptions, musical attributes, and optional reference audio, our system generates complete high-quality songs with controllable musical style and vocal characteristics.

The proposed framework is designed around the following key ideas:
\begin{itemize}
    \item \textbf{Semantic-aware RVQ tokenization and hierarchical modeling.} Our system uses an 8-codebook RVQ tokenizer to build a semantic-aware music representation, and hybrid-LM to capture both musical structure and fine details.

    \item \textbf{Full-song high-fidelity acoustic synthesis.} FullDiT performs non-causal flow matching in a continuous VAE latent space, conditioned on the complete 8-codebook codec sequence, lyrics, and text caption. This enables full-song acoustic rendering without relying on chunkwise synthesis.

    \item \textbf{Two-level melody modeling for cover generation.} The two-level melody module explicitly represents vocal melody with both coarse-grained and fine-grained tokens, improving melody preservation while retaining detailed pitch variation and singing expressiveness.

    \item \textbf{Preference alignment in post-training.} By combining DPO and GRPO for post-training, our system is further aligned with human preferences to improve musicality, controllability, and overall listening quality.
\end{itemize}

Empirically, we evaluate our system on a 500-example multilingual automatic benchmark. It obtains the highest point estimate in most reported automatic evaluation dimensions, while the Artificial Analysis Music with Vocals leaderboard assigns the submitted system an official rank range of 2--3. These results demonstrate competitive performance in the evaluated settings. Dedicated quantitative evaluation of instrumental and cover song generation remains future work. Overall, the proposed framework supports controllable, melody-aware, and high-fidelity song generation, bridging discrete music composition and full-length acoustic synthesis within a unified system.

%
%

Figure~\ref{fig:aa_leaderboard} provides an independent external reference from the Artificial Analysis Music with Vocals leaderboard~\cite{artificialanalysis2026music,artificialanalysis2026vocals}. Our submission occupies the third position by point estimate, with an Elo rating of 1,129 from 2,289 evaluation samples and an official rank range of 2--3. The gap to the second-ranked Mureka V8 is only 12 Elo points. Their reported 95\% confidence intervals, $[1{,}116, 1{,}144]$ and $[1{,}134, 1{,}150]$, overlap, and the point-estimate gap is small relative to the reported uncertainty. Accordingly, at the resolution of this leaderboard snapshot, the two systems occupy the same leading performance tier. This blind preference-based result complements the controlled automatic evaluation and suggests that our system's full-song vocal generation quality transfers to independent user judgments.

\FloatBarrier

\section{Method}

\subsection{Overview}

\begin{figure}[htbp]
    \centering
    \includegraphics[width=0.95\textwidth]{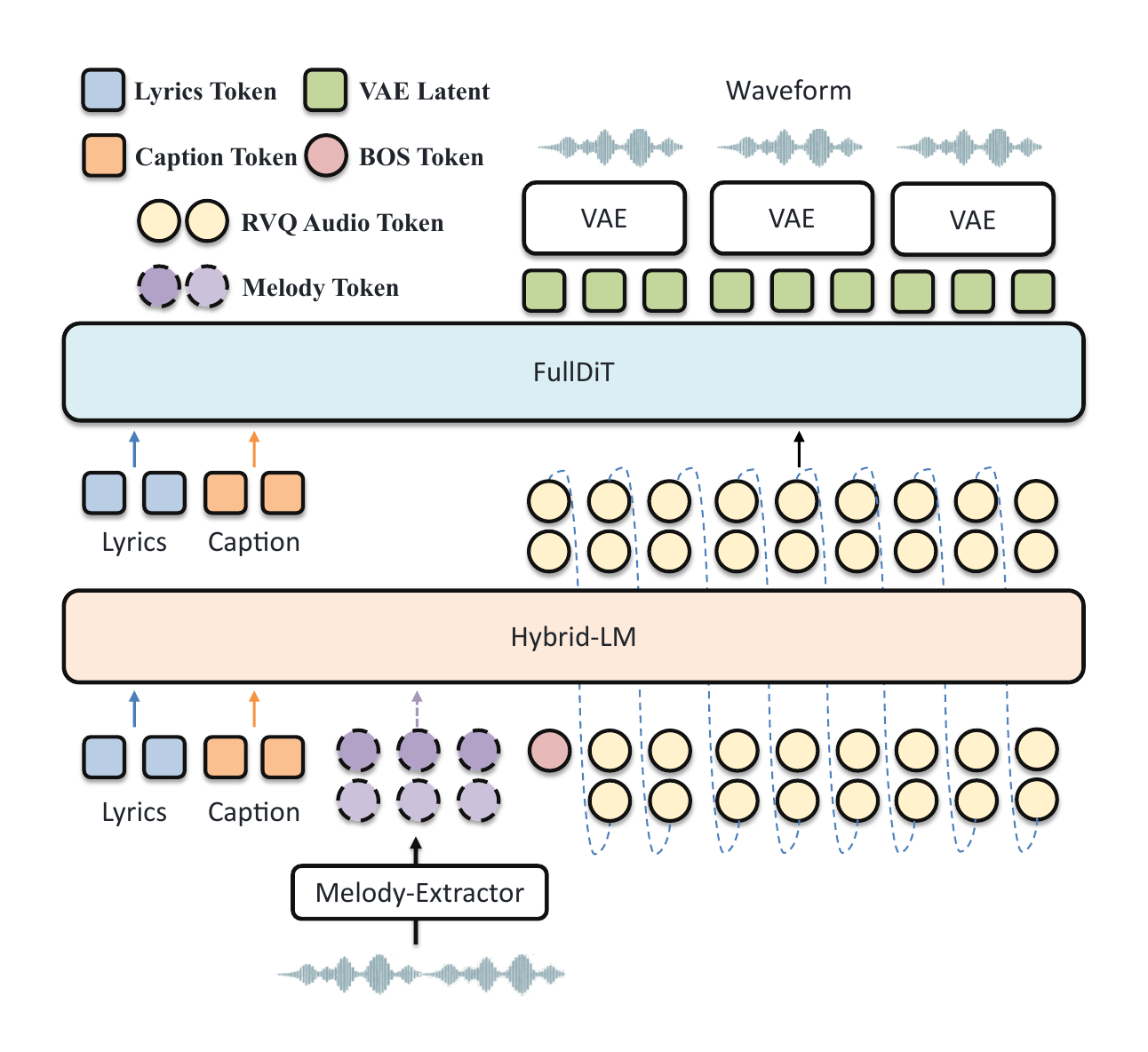}
    \caption{\textbf{Overview of the proposed framework.} Given a text caption, optional lyrics, and optional reference audio, hybrid-LM autoregressively generates audio tokens. For cover song generation, the melody module extracts and discretizes melody cues from the reference audio as additional conditions for hybrid-LM. FullDiT then generates a continuous full-song VAE latent conditioned on the codec sequence, text caption, and lyrics when available, which is decoded into a high-fidelity waveform.}

    \label{fig:overall}
\end{figure}

As shown in Figure~\ref{fig:overall}, the proposed framework transforms a natural-language music description, optional lyrics, and optional reference audio into a complete song. The system first constructs a structured textual condition from musical attributes and, when available, lyrics with section tags. Discrete audio tokens are then autoregressively predicted by hybrid-LM. For cover song generation, the melody module extracts and discretizes note-level and frame-level pitch cues from the reference audio and supplies them as additional conditions to hybrid-LM. Finally, FullDiT generates a continuous full-song VAE latent conditioned on the codec sequence, text caption, and lyrics when available, which is decoded into a high-fidelity waveform.

\begin{itemize}
    \item \textbf{Semantic-aware RVQ tokenizer} maps raw music audio into discrete token sequences using an 8-codebook residual vector quantization (RVQ) scheme. It is trained with a three-stage recipe consisting of BEST-RQ-style pretraining, multi-task finetuning, and discrete token training, ensuring that the resulting token stream preserves rich semantic information while maintaining high reconstruction fidelity.

    \item \textbf{hybrid-LM} generates audio tokens conditioned on lyrics, text descriptions, and musical attributes. It consists of an 8B-parameter global LLM and a 0.4B local LLM, which autoregressively predict the first-level audio tokens and the residual audio tokens, respectively. This design supports both vocal music and instrumental music generation.

    \item \textbf{Two-level melody module} extracts note-level and frame-level pitch cues from reference audio and discretizes them into coarse-grained and fine-grained melody tokens. These explicit melody conditions guide the overall vocal melody and local pitch details, helping preserve the core melody in cover song generation.

    \item \textbf{FullDiT} is an 8B-parameter DiT model that performs full-song flow matching in a continuous VAE latent space with non-causal self-attention. Conditioned on the codec sequence, text caption, and lyrics when available, it enriches acoustic detail and overall perceptual quality for high-fidelity full-song synthesis.
\end{itemize}

In the following sections, we first introduce the semantic-aware RVQ tokenizer and the discrete music representation it provides. We then describe hybrid-LM for autoregressive token generation, followed by FullDiT for high-fidelity song synthesis. Next, we present the two-level melody module for melody-guided cover song generation. Finally, we describe DPO, GRPO, and OPD as reward-based post-training strategies that incorporate automatically constructed preference signals, on-policy rewards, and teacher guidance to improve musicality.

\subsection{Semantic-Aware RVQ Tokenizer}

\begin{figure}[htbp]
    \centering
    \includegraphics[width=0.98\textwidth]{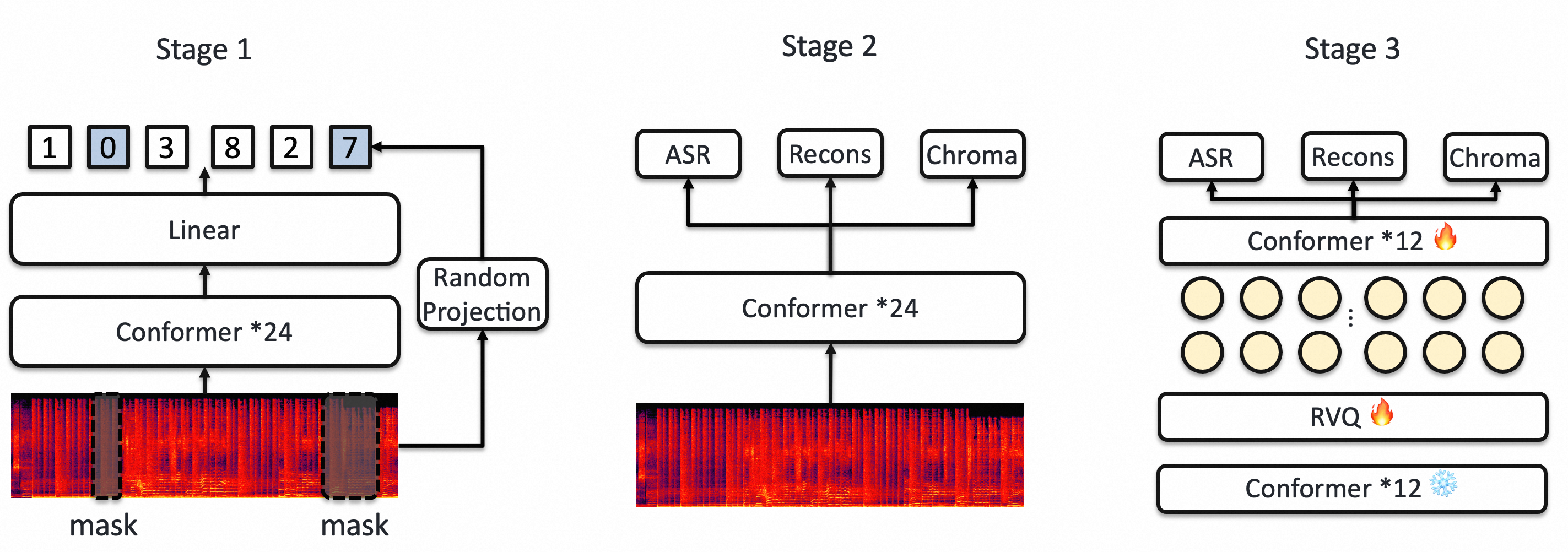}
    \caption{Three-stage training of the proposed tokenizer, which consists of BEST-RQ-style pretraining, multi-task finetuning, and RVQ token training.}
    \label{fig:tokenizer_train_stage}
\end{figure}

The proposed tokenizer maps raw music audio into discrete token sequences for downstream language modeling and acoustic generation. As shown in Figure~\ref{fig:tokenizer_train_stage}, we train it with a three-stage pipeline consisting of BEST-RQ-style pretraining\cite{chiu2022self}, multi-task finetuning, and RVQ token training.

\paragraph{Architecture.}
The tokenizer is built on a 24-layer Conformer encoder, followed by an 8-codebook residual vector quantization (RVQ) module. Each codebook has a size of 8192. Given input music audio features, the encoder produces continuous latent representations, which are then quantized into discrete token sequences by the RVQ module. The resulting tokens provide a compact yet expressive representation of musical content and serve as the target space for hybrid-LM.

\paragraph{Stage 1: BEST-RQ-style pretraining.}
In the first stage, we pretrain the 24-layer Conformer encoder with a BEST-RQ-style objective\cite{chiu2022self}. Masked audio features are encoded and trained to predict random-projection-quantized targets, enabling the encoder to learn robust contextual and semantic representations from large-scale music audio.

\paragraph{Stage 2: Multi-task finetuning.}
In the second stage, we finetune the encoder with multiple training objectives, including ASR, reconstruction, and chroma prediction, as shown in Figure~\ref{fig:tokenizer_train_stage}. This stage improves the representation quality for lyrics, acoustic details, and pitch-related musical structure.

\paragraph{Stage 3: RVQ token training.}
In the final stage, we introduce the RVQ module and train the tokenizer to generate discrete tokens. Specifically, we use the bottom 12 layers of the Conformer encoder as a frozen feature extractor, while the top 12 layers together with the 8-codebook RVQ module are optimized for tokenization. This stage yields discrete tokens that preserve semantic information while maintaining high reconstruction fidelity.

\subsection{hybrid-LM}

\begin{figure}[t]
    \centering
    \includegraphics[width=\textwidth]{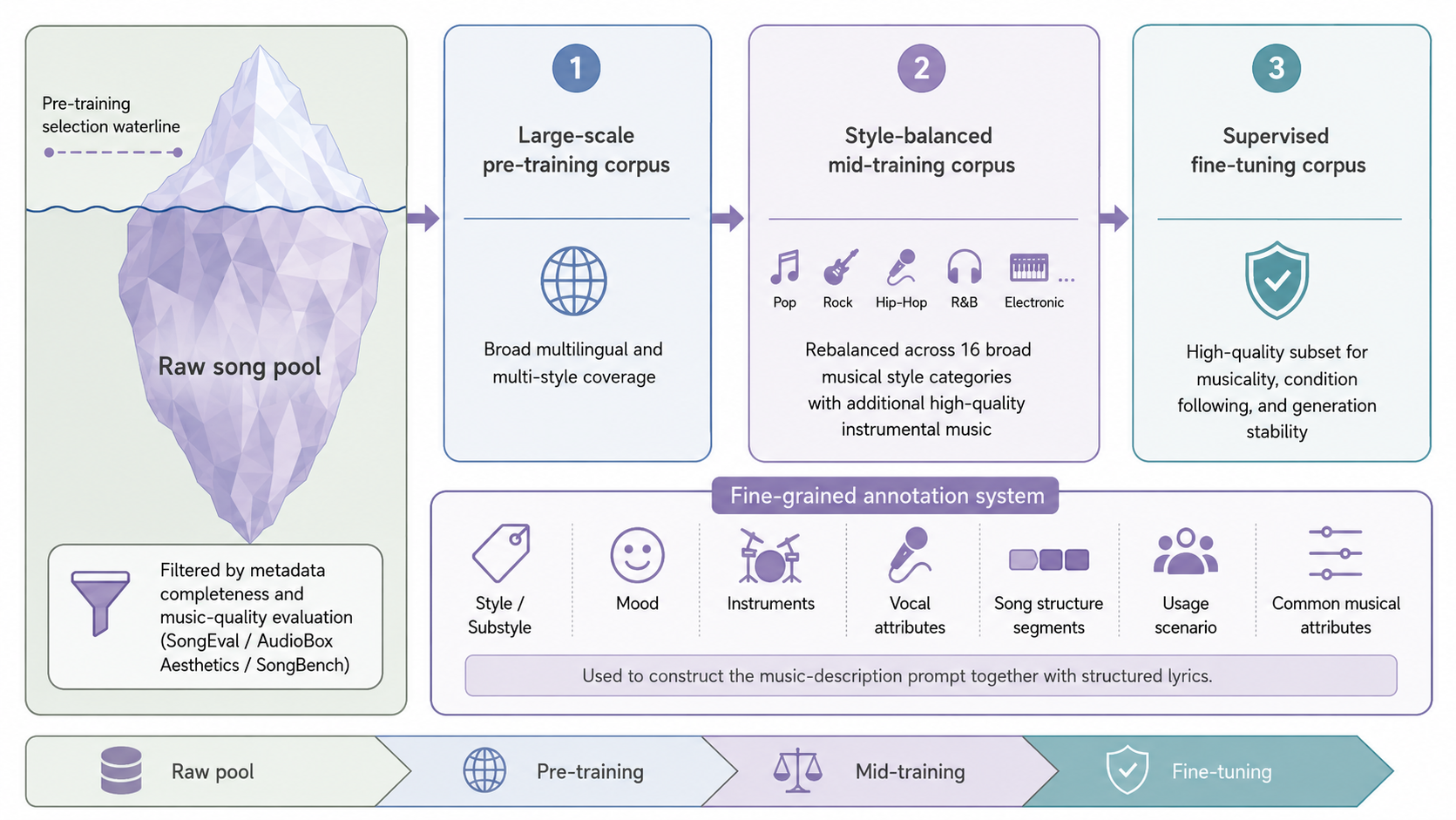}
    \caption{Overview of the music data curation and staged training pipeline.}
    \label{fig:data_pipeline}
\end{figure}

\subsubsection{Data and Training Pipeline}

The training of hybrid-LM follows a multi-stage pipeline consisting of large-scale pre-training, style-balanced mid-training, and high-quality supervised fine-tuning. We first collect a large-scale song corpus from diverse digital music sources, covering a wide range of languages, musical styles, vocal characteristics, and production practices. To construct a reliable training corpus, we apply a multi-stage curation process that considers metadata completeness together with complementary perceptual and musical-quality estimates from SongEval~\cite{yao2025songeval}, Audiobox Aesthetics~\cite{tjandra2025audiobox}, and SongBench~\cite{wu2026songbench}. After filtering and deduplication, a corpus containing tens of millions of songs is retained for large-scale training.

\paragraph{Fine-grained music annotation.}
An overview of the complete data curation and staged training pipeline is shown in Figure ~\ref{fig:data_pipeline}. To provide expressive and controllable conditioning signals, we design a fine-grained annotation system that describes each song from multiple complementary perspectives. The annotation taxonomy covers musical style and substyle, mood, instrumentation, vocal characteristics, segment-level song structure, usage scenario, and common musical attributes such as tempo, key, rhythmic characteristics, energy, and production-related descriptors. Structural annotations identify segments such as intro, verse, pre-chorus, chorus, bridge, instrumental break, and outro whenever applicable. These annotations are normalized into a structured representation and, together with lyrics and section markers, form the textual condition supplied to hybrid-LM.

\paragraph{Stage 1: Large-scale pre-training.}
The full curated corpus is used to establish broad multilingual and multi-style coverage. During this stage, the model learns general text-to-music correspondence, frame-level audio-token prediction, and long-range musical dependencies. We pre-train hybrid-LM using a relatively high learning rate for a limited number of passes over the large-scale corpus.

\paragraph{Stage 2: Style-balanced mid-training.}
Large-scale music corpora naturally exhibit a long-tailed distribution in which a small number of popular styles dominate the training data. We therefore use the SongBench Musicality score to select a higher-quality subset and rebalance it across a broad set of musical style categories, including Pop, Rock, Folk, Hip-Hop, R\&B, Soul, and EDM, etc. Together with additional high-quality instrumental music, this process yields a multi-million-song training corpus. We continue training on this corpus using a reduced learning rate. This stage improves both musical quality and style coverage while reducing over-specialization toward dominant genres.

\paragraph{Stage 3: Supervised fine-tuning.}
For the final stage, we select a compact subset of high-quality songs from the curated data pool. The selected examples exhibit strong musicality, reliable annotations, clear structural organization, and close agreement between the audio and its textual conditions. We perform supervised fine-tuning using a substantially lower learning rate. This final refinement stage further improves musical coherence, condition following, generation stability, and overall perceptual quality.

\begin{figure}[h]
    \centering
    \includegraphics[width=0.9\textwidth]{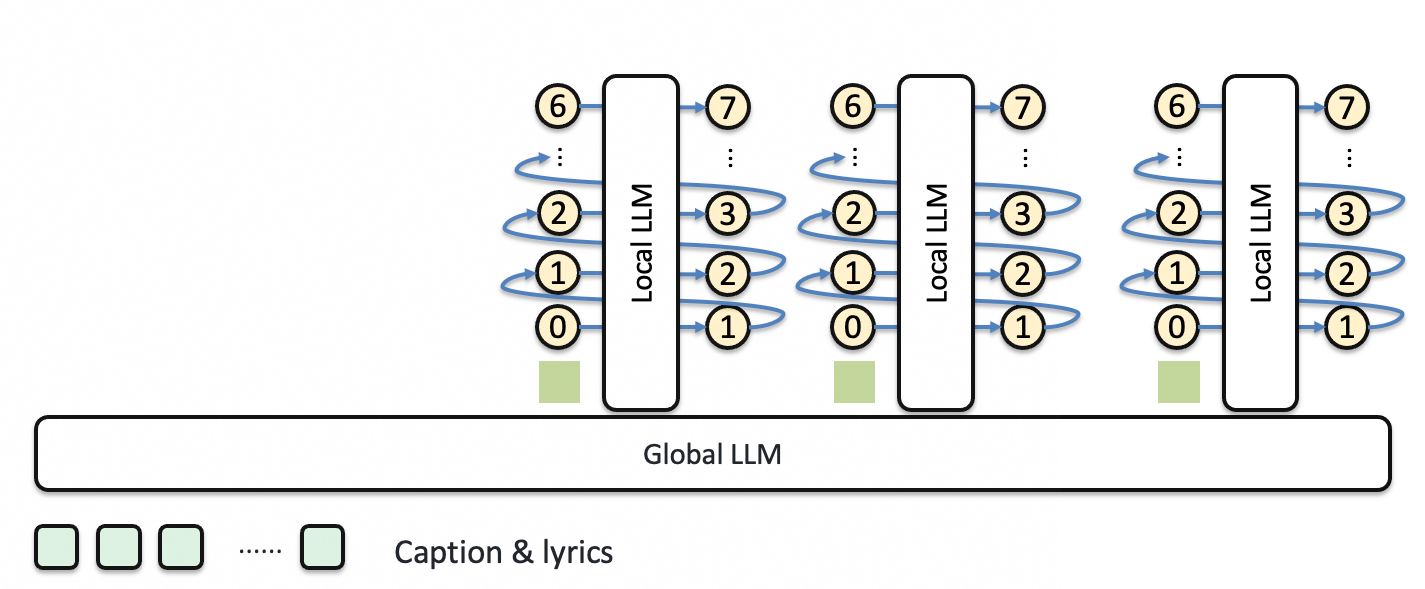}
    \caption{Overview of the hierarchical autoregressive architecture of hybrid-LM. Structured music annotations and lyrics are provided as textual conditions, with optional frame-aligned melody tokens used for cover generation. The 8B global LLM predicts the first codebook token $c_t^0$ once per audio frame and models long-range temporal dependencies, while the 0.4B local LLM autoregressively predicts the residual codebook tokens $c_t^1$--$c_t^7$ within the frame.}
    \label{fig:lm_model}
\end{figure}

\subsubsection{Hierarchical Autoregressive Architecture}

As illustrated in Figure ~\ref{fig:lm_model}, hybrid-LM adopts a hierarchical autoregressive architecture that separates long-range temporal modeling from within-frame acoustic modeling. Following the factorization introduced by HeartMuLa~\cite{yang2026heartmula}, the architecture consists of an 8B-parameter global LLM operating along the temporal axis and a lightweight 0.4B-parameter local LLM operating along the RVQ codebook axis.

The model receives structured music annotations and lyrics as textual conditions and predicts discrete audio tokens generated by the proposed tokenizer. Let $X$ denote the textual condition, comprising the structured music description and lyrics, and let $M=(M^{\mathrm{MIDI}},M^{\mathrm{F0}})$ denote the optional frame-aligned melody condition used for cover generation. For non-cover generation, $M=\emptyset$. Given an audio-token sequence $A=(\mathbf{c}_1,\ldots,\mathbf{c}_T)$, each audio frame is represented by eight RVQ tokens, $\mathbf{c}_t=(c_t^0,\ldots,c_t^7)$. The first codebook $c_t^0$ provides a compact frame-level representation for global temporal modeling, while $c_t^1$--$c_t^7$ progressively encode residual acoustic information. Each codebook contains 8,192 entries together with an end-of-sequence token, and the tokenizer operates at a frame rate of 25~Hz.

We factorize the conditional audio-token distribution as
\begin{equation}
p(A\mid X,M)=\prod_{t=1}^{T}\left[p\!\left(c_t^0\mid\mathbf{c}_{<t},X,M;\theta_{\mathrm{g}}\right)\prod_{k=1}^{7}p\!\left(c_t^k\mid h_t^{\mathrm{g}},\mathbf{c}_t^{<k};\theta_{\mathrm{l}}\right)\right],
\label{eq:lm_hierarchical_factorization}
\end{equation}
where $h_t^{\mathrm{g}}$ denotes the global hidden state conditioned on $X$ and $M$, $\mathbf{c}_{<t}$ represents all preceding audio frames, and $\mathbf{c}_t^{<k}$ denotes the codebooks already generated within the current frame. The parameters of the global and local models are denoted by $\theta_{\mathrm{g}}$ and $\theta_{\mathrm{l}}$, respectively.

Under this factorization, the large global LLM is evaluated only once per audio frame, while the lightweight local LLM completes the remaining seven codebooks along a short, fixed-length depth axis. Compared with flattening all eight codebooks into a single temporal token sequence, this design substantially reduces the sequence length and computational workload of the global model while retaining the acoustic capacity provided by multi-codebook representations.

\paragraph{Global LLM.}
The global component is initialized from Qwen3-8B and serves as the temporal backbone of hybrid-LM. Structured music annotations and lyrics are serialized as a text prefix, while the embeddings of the eight codebooks from each preceding audio frame are aggregated into a single frame-level representation. At each temporal step, the global LLM predicts the first codebook token $c_t^0$ and produces the hidden state $h_t^{\mathrm{g}}$ for the local branch. Operating at the frame level allows the model to focus on long-range musical dependencies, including melody development, rhythmic continuity, lyric progression, section transitions, repetition, and overall song structure. Its extended context window supports complete-song generation of up to five minutes.

\paragraph{Local LLM.}
The local component is a lightweight 0.4B-parameter causal Transformer trained from scratch. Conditioned on the global hidden state $h_t^{\mathrm{g}}$ and the first codebook token $c_t^0$, it autoregressively predicts the remaining codebooks $c_t^1$ through $c_t^7$ along the codebook axis. Teacher forcing is used during training, while the residual codebooks are generated sequentially during inference. Once all eight codebooks of frame $t$ have been generated, their embeddings are aggregated into a frame-level representation and fed back to the global LLM for prediction of the next frame.

\paragraph{Prompt conditioning and classifier-free guidance.}
The text prompt consists of two complementary components: a music description and structured lyrics. The music description specifies high-level attributes such as style, mood, instrumentation, vocal characteristics, usage scenario, and other common musical properties. The lyrics provide the textual content of the song and retain explicit section markers, such as \texttt{[Verse]}, \texttt{[Chorus]}, \texttt{[Bridge]}, and \texttt{[Instrumental]}, to describe the intended song structure. Together, these two components provide control over the overall musical characteristics, lyrical content, and sectional organization of the generated song.

To enable classifier-free guidance, the textual condition is randomly removed for 10\% of the training examples and replaced with a learned unconditional embedding. During inference, the conditional and unconditional predictions can be combined to strengthen adherence to the requested musical attributes and lyrics without changing the hierarchical decoding procedure.

\paragraph{Weighted training objective.}
The global and local components are jointly optimized using masked next-token cross-entropy. Let $\mathcal{M}_k$ denote the set of valid target positions associated with codebook $k$. The loss for each codebook is defined as
\begin{equation}
\mathcal{L}_k=-\frac{1}{|\mathcal{M}_k|}\sum_{t\in\mathcal{M}_k}\log p\!\left(c_t^k\mid X,\mathbf{c}_{<t},\mathbf{c}_t^{<k}\right).
\label{eq:lm_codebook_loss}
\end{equation}

Because the seven residual codebooks represent related acoustic refinements, directly summing their losses would cause the local branch to dominate the total objective merely because it contains more prediction targets. We therefore average the residual-codebook losses and assign a larger weight to the first codebook:
\begin{equation}
\mathcal{L}_{\mathrm{audio}}=5\mathcal{L}_0+\frac{1}{7}\sum_{k=1}^{7}\mathcal{L}_k.
\label{eq:lm_weighted_audio_loss}
\end{equation}

The global $c^0$ objective and the aggregated local objective therefore have a weight ratio of $5{:}1$. This weighting reflects the asymmetric roles of the two branches: errors in $c^0$ directly affect temporal planning and propagate to subsequent frames, whereas $c^1$--$c^7$ primarily refine timbre and local acoustic detail. The textual prefix is used solely as conditioning context and is excluded from the audio-token prediction loss.

\subsection{FullDiT}

  \label{subsec:fulldit}

  \begin{figure}[t]
      \centering
      \includegraphics[width=0.9\textwidth]{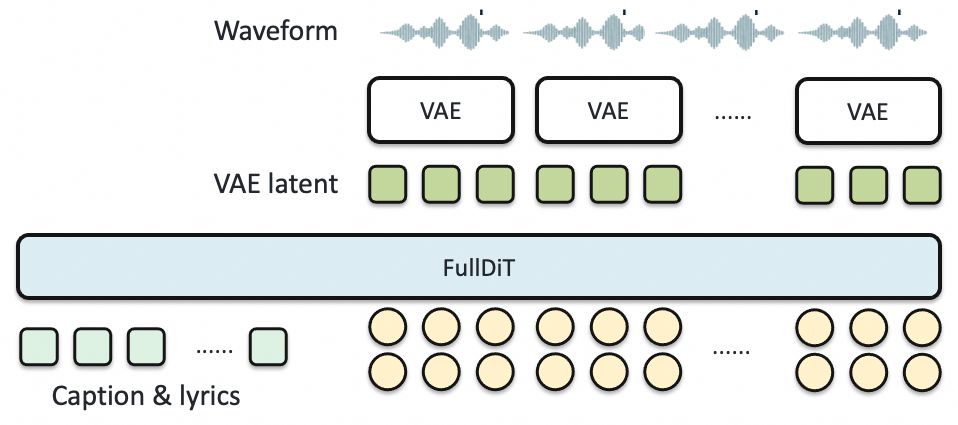}
      \caption{\textbf{Architecture of FullDiT.}
      Given full-song RVQ tokens, lyrics, and a text caption,
      FullDiT injects frame-aligned codec embeddings into
      the audio hidden states and performs non-causal full-song flow
      matching to generate a continuous VAE latent.}
      \label{fig:fm_model}
  \end{figure}

  FullDiT is an 8B-parameter conditional DiT that converts the
  discrete music plan generated by hybrid-LM into a high-fidelity
  continuous acoustic representation. As illustrated in Figure ~\ref{fig:fm_model}, it takes the full sequence of
  8-codebook RVQ tokens together with the original lyrics and text caption,
  and generates an acoustic latent that is subsequently converted into a
  48-kHz stereo waveform by a fixed VAE decoder. Instead of treating the
  audio tokens as a deterministic reconstruction target, the model performs
  conditional acoustic generation from Gaussian noise, allowing it to enrich
  acoustic details and correct locally imperfect token predictions.

  \paragraph{High-fidelity data for FullDiT.}
  Starting from the pre-training pool, we construct a separate renderer corpus with stricter acoustic-quality constraints. The selection jointly considers effective bandwidth, stereo image width, Audiobox quality estimates, and source bitrate in order to reject bandwidth-limited, spatially collapsed, or otherwise degraded audio. The resulting subset of lossless audio is used to train FullDiT, enabling the renderer to synthesize high-fidelity music in 48-kHz stereo.

  \paragraph{Frame-aligned codec conditioning.}
  For each time frame, the eight RVQ tokens are mapped through
  codebook-specific embedding tables, fused, and projected to the hidden
  dimension of the DiT. The resulting codec condition is directly added to
  the audio state at the corresponding time frame:
  \begin{equation}
  e^{\mathrm{codec}}_i
  =
  \operatorname{Fuse}_{k=1}^{8}
  \operatorname{Emb}_k(c_{i,k}),
  \qquad
  a^{(0)}_i
  =
  \operatorname{EmbedAudio}(z_{t,i})
  +
  e^{\mathrm{codec}}_i ,
  \label{eq:fulldit_codec_condition}
  \end{equation}
  where $c_{i,k}$ denotes the token from codebook $k$ at frame $i$, and
  $z_{t,i}$ denotes the corresponding frame of the noisy VAE latent.
  This design preserves the temporal correspondence between the discrete
  music plan and the generated acoustic representation.

  \paragraph{Full-context text and audio modeling.}
  Lyrics and captions are independently encoded and provided directly to
  FullDiT rather than being used only by hybrid-LM.
  The caption represents global attributes such as genre, instrumentation,
  mood, and production style, while the lyrics provide phonetic content and
  section-level structure. In each DiT block, the audio states serve as
  queries, while the encoded lyrics, caption, and audio states jointly form
  the key-value context. In addition, the audio sequence is modeled with
  non-causal self-attention over the entire song. The model can process up to
  8192 frames, corresponding to 327.68 seconds at 25 Hz, enabling it to model
  verse--chorus repetition, section transitions, recurring motifs, and
  long-range timbral consistency within a single generation trajectory.

    \paragraph{Flow-matching objective.}
    Let $\sigma\in[0,1]$ denote the flow-path coordinate, with $\sigma=0$ corresponding to data and $\sigma=1$ to Gaussian noise. Given a target acoustic latent $z_0$ and $\epsilon\sim\mathcal{N}(0,I)$, we define
    \begin{equation}
    z_{\sigma}=(1-\sigma)z_0+\sigma\epsilon,
    \qquad
    v_{\sigma}^{*}=\epsilon-z_0.
    \label{eq:fulldit_flow_path}
    \end{equation}
    %
    Conditioned on the codec sequence $c$, lyrics $l$, and caption $p$, FullDiT is trained with the following flow-matching objective:
    \begin{equation}
    \mathcal{L}_{\mathrm{FM}}
    =
    \mathbb{E}
    \left[
    w(\sigma)
    \left\|
    v_{\theta}(z_{\sigma},\sigma,c,l,p)
    -
    (\epsilon-z_0)
    \right\|_2^2
    \right].
    \label{eq:fulldit_flow_loss}
    \end{equation}
    During inference, the corresponding reverse trajectory is solved from Gaussian noise to the generated acoustic latent, which is then decoded into the final waveform.

  \paragraph{Error-Matched Distractor Conditioning.}
  During training, codec tokens extracted from the target audio are perfectly
  aligned with the target latent, whereas inference-time tokens predicted by
  hybrid-LM may contain local errors. To reduce this train--inference
  mismatch, we introduce Error-Matched Distractor Conditioning (EMDC).
  For each codebook $k$, a token is replaced with probability
  $p_k=1-\mathrm{Acc@1}_k$, where $\mathrm{Acc@1}_k$ is the teacher-forced
  top-1 accuracy of the upstream hybrid-LM. When replacement is applied,
  a near-miss token is sampled from the cosine top-$K_k$ neighborhood of the
  original token in the corresponding codec embedding space. Importantly,
  EMDC modifies only the codec condition while keeping the original acoustic
  target $z_0$ unchanged. The model therefore learns to selectively correct
  plausible codec errors using the remaining codec information, text
  conditions, full-song context, and its learned acoustic prior.

  \paragraph{Four-way classifier-free guidance.}
  At inference time, we use four-way classifier-free guidance to control the
  contributions of codec, lyrics, and caption independently. Let
  $v_u$, $v_k$, $v_{kl}$, and $v_{klc}$ denote the vector-field predictions
  from the unconditional, codec-only, codec-plus-lyrics, and fully
  conditioned branches, respectively. The final prediction is
  \begin{equation}
  \begin{aligned}
  v_{\mathrm{cfg}}
  ={}&
  v_u
  +s_{\mathrm{codec}}(v_k-v_u) \\
  &+s_{\mathrm{lyrics}}(v_{kl}-v_k)
  +s_{\mathrm{caption}}(v_{klc}-v_{kl}),
  \end{aligned}
  \label{eq:fulldit_cfg}
  \end{equation}
  where the three guidance scales independently control adherence to the
  audio-token plan, lyrics, and global musical description. This
  decomposition provides flexible control over content fidelity and acoustic
  generation quality without requiring separate models.

\subsection{Two-Level Melody Modeling}

\begin{figure}[htbp]
    \centering
    \includegraphics[width=0.7\textwidth]{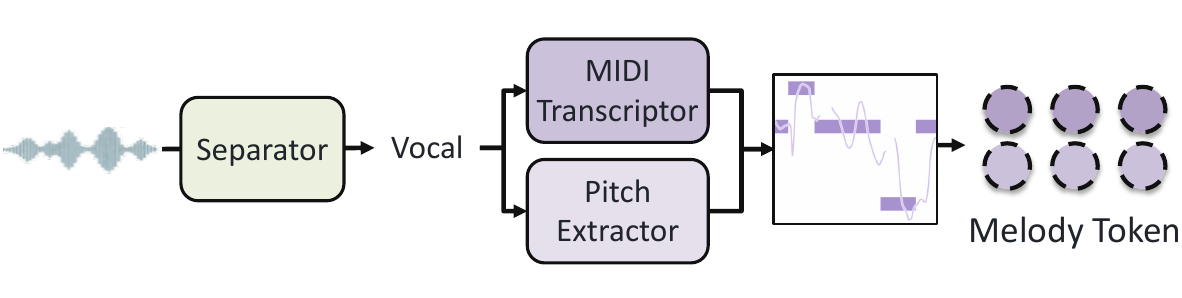}
    \caption{The proposed two-level melody tokenizer.}
    \label{fig:melody_tokenizer}
\end{figure}

The proposed melody module models vocal melody with two levels of melodic features, including coarse-grained note-level features and fine-grained pitch-level features. This design provides explicit melodic guidance for cover song generation, allowing the model to capture both the global contour of the singing melody and local pitch variations.

\paragraph{Melody feature extraction.}
The architecture of the two-level melody tokenizer is illustrated in Figure~\ref{fig:melody_tokenizer}. Given a raw input audio, we first apply a BS-RoFormer separator \cite{lu2024music} to extract the vocal track from the mixture. Based on the separated vocal track, we then extract two complementary melody features. First, a MIDI transcription model is used to estimate vocal note sequences, which serve as coarse-grained melodic features. These note-level features describe the overall contour and progression of the singing melody. Second, a pitch extraction model is applied to estimate frame-level F0 values, which serve as fine-grained melodic features. Compared with MIDI notes, F0 captures more detailed local pitch variations and expressive singing dynamics.

\paragraph{Melody tokenization.}
After feature extraction, we convert both note-level and pitch-level representations into frame-aligned discrete token sequences at 25~Hz. For MIDI notes, each note is repeated along the temporal axis according to its duration, and the corresponding MIDI pitch ID is used as a 128-level discrete token. The MIDI ID of 0 is reserved for non-vocal frames. For F0, we first transform it into the log-pitch domain so that it is better aligned with the pitch scale of MIDI notes. We then quantize it into a 128-level discrete token sequence at 25~Hz, where the zero value is similarly reserved for non-vocal frames.

The resulting MIDI tokens and F0 tokens are fed into hybrid-LM in parallel, forming a two-codebook discrete melody representation. The coarse-grained MIDI tokens guide the overall vocal melody and help preserve the main melodic contour of the reference song, while the fine-grained F0 tokens model local pitch details, enabling more accurate vocal realization and lyric-to-pitch alignment.

\subsection{Reward-Based Post-Training}
\begin{figure}[htbp]
    \centering
    \includegraphics[width=0.9\textwidth]{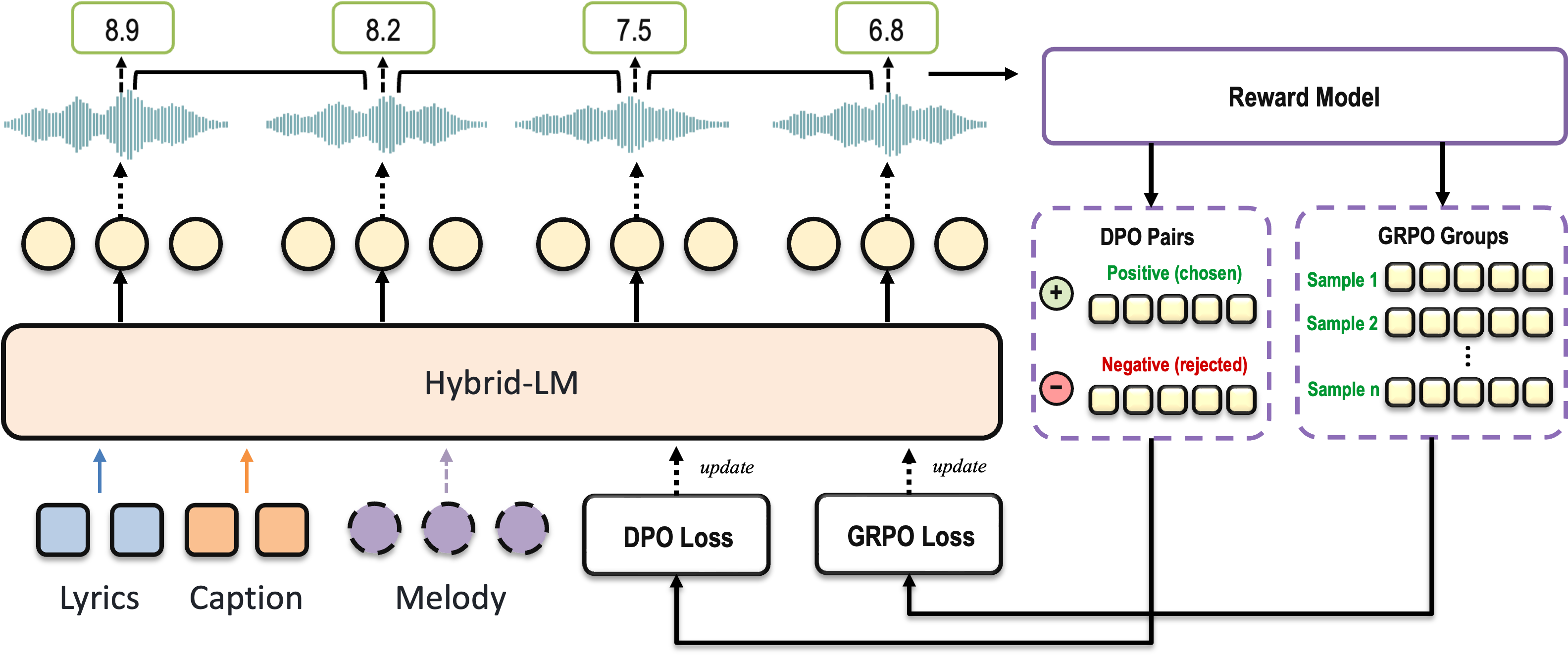}
    \caption{Overview of the RL-based post-training pipeline, including DPO and GRPO.}
    \label{fig:dpo}
\end{figure}

Although large-scale pre-training enables the model to acquire broad music-generation capabilities, the pre-training corpus inevitably contains samples of varying quality. Such data heterogeneity may lead to unstable generation performance, particularly in terms of musical structure, melodic coherence, rhythmic consistency, harmonic plausibility, and overall audio quality. To further improve the lower bound of generation quality and reduce performance variance across different prompts, we introduce a dedicated post-training stage for hybrid-LM consisting of
supervised fine-tuning (SFT), Direct Preference Optimization (DPO), Group Relative Policy Optimization (GRPO), and On-Policy Distillation (OPD). DPO learns from offline preference pairs, whereas GRPO directly optimizes rewards on samples generated by the current policy. OPD further complements the sparse sequence-level reward with dense token-level guidance from a frozen high-quality teacher. The illustration of DPO and GRPO are shown in Figure~\ref{fig:dpo}.
These post-training procedures primarily focus on improving the musicality of the generated results while preserving the general generation capability acquired during pre-training. Beyond hybrid-LM, we additionally apply GRPO-based post-training to FullDiT to refine audio fidelity at the waveform level, targeting improved vocal-accompaniment balance and stereo spatial rendering while keeping the pre-trained generation diversity intact.

\subsubsection{Supervised Fine-Tuning}
\label{subsec:sft}

As described in Stage 3 of the hybrid-LM training pipeline, supervised fine-tuning is performed on a compact, carefully curated subset of high-quality songs. We optimize the model on this subset using the standard supervised generation objective:
\begin{equation}
\mathcal{L}_{\mathrm{SFT}}
=
-\mathbb{E}_{(x,y)\sim\mathcal{D}_{\mathrm{SFT}}}
\left[
\log \pi_{\theta}(y \mid x)
\right],
\label{eq:sft_loss}
\end{equation}
where $x$ denotes the conditioning prompt, $y$ denotes the corresponding 
high-quality music sample, and $\pi_{\theta}$ represents the trainable music 
generation model. This process shifts the model distribution toward 
higher-quality musical samples and provides a stable initialization for 
subsequent preference optimization.

\subsubsection{Direct Preference Optimization}
\label{subsec:dpo}

To construct the preference dataset for DPO training, the SFT model independently
generates multiple candidate songs for each prompt $x$ under the same generation
configuration. The candidates are evaluated using SongBench, which assigns an automatic score $s(y_i)$ to each generated sample according to its overall musical quality.
Candidate pairs are then constructed according to the difference between the scores.

For two candidates $y_i$ and $y_j$, a preference pair is retained only when 
their score difference exceeds a predefined threshold $\delta$. The 
higher-scoring candidate is treated as the preferred sample $y^{+}$, whereas 
the lower-scoring candidate is treated as the rejected sample $y^{-}$. The resulting preference dataset is defined as
\begin{equation}
\mathcal{D}_{\mathrm{DPO}}
=
\left\{
\left(x, y^{+}, y^{-}\right)
\,\middle|\,
s(y^{+}) - s(y^{-}) > \delta
\right\}.
\label{eq:dpo_dataset}
\end{equation}
This threshold-based pairing strategy filters out ambiguous comparisons in 
which two candidates exhibit similar levels of musical quality. It therefore 
reduces noise in the automatically constructed preference labels and ensures 
that the model is trained on sufficiently distinguishable positive and 
negative examples.

Given a preference triplet $(x,y^{+},y^{-})$, DPO encourages the trainable model 
to assign a higher relative likelihood to the preferred sample than to the 
rejected sample. The DPO objective is formulated as

\begin{equation}
r_{\theta}(x,y)
=
\beta
\log
\frac{\pi_{\theta}(y\mid x)}
     {\pi_{\mathrm{ref}}(y\mid x)}.
\label{eq:implicit_reward}
\end{equation}

\begin{equation}
\mathcal{L}_{\mathrm{DPO}}
=
-\mathbb{E}_{(x,y^{+},y^{-})\sim\mathcal{D}_{\mathrm{DPO}}}
\left[
\log \sigma
\left(
r_{\theta}(x,y^{+})-r_{\theta}(x,y^{-})
\right)
\right].
\label{eq:dpo_loss}
\end{equation}

where $r_{\theta}(x,y)$ denotes the implicit reward defined by the 
log-likelihood ratio between the trainable policy $\pi_{\theta}$ and 
the frozen reference policy. The DPO pipeline is illustrated in Figure~\ref{fig:dpo}.

In addition to the standard DPO objective, we introduce an auxiliary SFT loss 
computed only on the preferred samples. Thus, the final optimization objective is defined as
\begin{equation}
\mathcal{L}_{\mathrm{post}}
=
\mathcal{L}_{\mathrm{DPO}}
+
\lambda_{\mathrm{SFT}}
\mathcal{L}_{\mathrm{SFT}}^{+},
\label{eq:joint_dpo_sft_loss}
\end{equation}
where $\lambda_{\mathrm{SFT}}$ controls the contribution of the auxiliary 
positive-sample SFT objective.

The standard DPO loss optimizes the relative preference margin between the 
preferred and rejected generations, whereas the auxiliary SFT loss directly 
increases the likelihood of the preferred samples. Combining the two objectives 
prevents the model from relying exclusively on relative ranking signals and 
provides an explicit learning signal toward high-quality musical outputs. This 
joint training objective improves optimization stability, preserves the 
fundamental generation capabilities learned during pre-training and SFT, and 
encourages the output distribution to move consistently toward samples with 
stronger musicality.

\subsubsection{Group Relative Policy Optimization}
\label{sec:hybrid-lm-grpo}

For each conditioning input $x$, which consists of a caption, lyrics, and
optional melody conditions, hybrid-LM independently samples a group of $G$
candidate music-token sequences $\{y_i\}_{i=1}^{G}$ under the same generation
configuration. Each candidate is rendered into audio and automatically
evaluated by SongBench for overall musical quality. SongBench scores melody,
arrangement, structure, likability, vocal quality, instrumental quality, and
mixing, and aggregates these dimensions into a scalar reward $r_i$.

For each group, we compute the normalized relative advantage
\begin{equation}
A_i
=
\frac{r_i - \mu_G}{\sigma_G + \epsilon},
\label{eq:grpo_advantage}
\end{equation}
where $\mu_G$ and $\sigma_G$ denote the mean and standard deviation of rewards
within the group. To maintain a reliable and stable reward signal, groups whose
reward variance falls below a minimum threshold are excluded from the
policy-gradient objective.

Let
\begin{equation}
\rho_{i,t}(\theta)
=
\frac{
\pi_{\theta}(y_{i,t}\mid x,y_{i,<t})
}{
\pi_{\mathrm{old}}(y_{i,t}\mid x,y_{i,<t})
},
\label{eq:grpo_ratio}
\end{equation}
be the token-level importance ratio. The clipped GRPO objective is
\begin{equation}
\mathcal{L}_{\mathrm{GRPO}}
=
-\mathbb{E}_{i,t}
\left[
\min\left(
\rho_{i,t} A_i,
\operatorname{clip}
\left(
\rho_{i,t},
1-\epsilon_l,
1+\epsilon_h
\right) A_i
\right)
\right]
+
\beta\mathcal{L}_{\mathrm{ref\text{-}KL}},
\label{eq:grpo_loss}
\end{equation}
where the reference-policy KL term constrains excessive deviation from the
initialization policy.

Optimizing solely with the group-relative policy objective can be unstable,
especially for long music sequences in which thousands of tokens share a
single terminal reward. To stabilize training and provide a denser learning
signal, we augment GRPO with an auxiliary online rejection-sampling fine-tuning
(RSFT) objective. For each on-policy group, we select the highest-reward sample
as a pseudo-label and optimize its music tokens using standard token-level
negative log-likelihood, denoted by $\mathcal{L}_{\mathrm{RSFT}}$.
The resulting GRPO--RSFT objective is
\begin{equation}
\mathcal{L}_{\mathrm{GRPO\text{-}RSFT}}
=
\mathcal{L}_{\mathrm{GRPO}}
+
\lambda_{\mathrm{RSFT}}
\mathcal{L}_{\mathrm{RSFT}},
\label{eq:grpo_rsft_loss}
\end{equation}
Here, $\lambda_{\mathrm{RSFT}}$ controls the contribution of the auxiliary
RSFT objective, while GRPO remains the primary policy-optimization objective.

Overall, GRPO alignment improves musicality, global structure, and
listenability, resulting in more coherent and stable full-song generation.

\subsubsection{On-Policy Distillation}
\label{sec:hybrid-lm-opd}

The GRPO stage improves sequence-level music quality through on-policy
exploration, but its scalar terminal reward provides limited token-level
guidance. Although online RSFT supplies a denser signal, it retains only one
sampled sequence per group and converts the reward ranking into a one-hot token
target. Meanwhile, an SFT teacher trained on curated high-quality data captures
a strong prior over high-quality music generation, but is slightly overfitted
to this narrower data distribution and consequently exhibits reduced generation
diversity. Following GRPO training, we introduce On-Policy Distillation (OPD)
while retaining the GRPO objective. The student continues to sample trajectories
from its current policy, preserving the quality gains from GRPO and the
student's generation diversity, while the frozen teacher provides soft
distributional guidance on the same student-generated prefixes.
In this way, OPD transfers the teacher's high-quality prior without requiring
direct imitation of teacher-generated trajectories or overly constraining
policy exploration.

The student first generates trajectories with its current policy. For every
valid response-token position, the same sampled prefix is then fed to both the
frozen teacher and the trainable student under teacher forcing. Let
$p_{\mathrm{tea}}^{(T)}$ and $p_{\theta}^{(T)}$ denote their
temperature-scaled full-vocabulary distributions conditioned on the same
sampled prefix. The OPD loss is
\begin{equation}
\mathcal{L}_{\mathrm{OPD}}
=
\frac{T^2}{|M|}
\sum_{t\in M}
D_{\mathrm{KL}}
\left(
p_{\mathrm{tea}}^{(T)}(\cdot\mid x,y_{<t})
\parallel
p_{\theta}^{(T)}(\cdot\mid x,y_{<t})
\right),
\label{eq:opd_loss}
\end{equation}
where $M$ is the set of valid response-token positions and $T$ is the
distillation temperature. The combined GRPO--OPD objective is
\begin{equation}
\mathcal{L}_{\mathrm{GRPO\text{-}OPD}}
=
\mathcal{L}_{\mathrm{GRPO}}
+
\lambda_{\mathrm{OPD}}
\mathcal{L}_{\mathrm{OPD}},
\label{eq:grpo_opd_loss}
\end{equation}

The teacher remains fixed throughout distillation. A moderate OPD weight
regularizes the student without suppressing exploration, whereas an overly
large weight can force rapid convergence toward the teacher and weaken the
benefit of online reward optimization. OPD improves generation quality and
stylistic consistency by transferring the teacher's high-quality prior, while
retaining the diversity of the on-policy student distribution.

\subsubsection{Post-Training for FullDiT}
\label{subsec:flow_grpo}


We apply FlowTTS-GRPO~\cite{flowtts_grpo, flowse_grpo, flowgrpo} to FullDiT for post-training refinement, targeting audio fidelity and musical balance. We convert deterministic Ordinary Differential Equation (ODE) sampling into a Stochastic Differential Equation (SDE) sampling for on-policy exploration and optimize group-relative terminal rewards. The policy model is conditioned on caption, lyrics, and discrete tokens. The reward combines ViSQOL\cite{hines2015visqol, chinen2020visqol} audio quality and an internal music reward model, normalized by their per-batch standard deviations:

\begin{equation} R = \lambda_1 \frac{R_{\mathrm{ViSQOL}}}{\mathrm{std}(R_{\mathrm{ViSQOL}})} + \lambda_2 \frac{R_{\mathrm{Music}}}{\mathrm{std}(R_{\mathrm{Music}})}, \end{equation}

This normalization makes $\lambda_1$ and $\lambda_2$ express the intended objective balance rather than raw reward variance. This post-training stage further improves vocal-accompaniment balance and stereo spatial rendering, encouraging the model to produce well-balanced mixes with immersive spatial presentation. SDE exploration is restricted to an early-step window, and classifier-free guidance~\cite{DBLP:journals/corr/abs-2207-12598} is omitted during rollouts.

\section{Results}

  \subsection{Controlled FullDiT Ablations}

  To isolate the contributions of the FullDiT design choices from model scale,
  we additionally train a complete 1.5B-parameter FullDiT (M1) specifically for
  controlled ablation studies. We construct three matched 1.5B variants that
  remove EMDC (M2), full-song training context (M3a), or decoder-side text
  conditioning (M3b), respectively. All comparisons in this subsection are
  therefore conducted among the 1.5B models in Table~\ref{tab:fulldit_ablation_models};
  the end-to-end system results reported elsewhere use the 8B FullDiT described
  in Section~\ref{subsec:fulldit}.

  \begin{table}[t]
      \centering
      \caption{Configurations of the controlled 1.5B FullDiT models. Except for
      the factor identified in the final column, the ablation models follow the
      same architecture and supervised-training setup as M1.}
      \label{tab:fulldit_ablation_models}
      \begingroup
      \scriptsize
      \setlength{\tabcolsep}{2.5pt}
      \renewcommand{\arraystretch}{1.08}
      \begin{tabular*}{\linewidth}{@{\extracolsep{\fill}}lcclll@{}}
          \toprule
          \textbf{Model} &
          \textbf{Scale} &
          \textbf{EMDC} &
          \textbf{Training audio context} &
          \textbf{Decoder-side text} &
          \textbf{Role} \\
          \midrule
          M1  & 1.5B & on  & full song & caption + lyrics & complete controlled
          model \\
          M2  & 1.5B & off & full song, & caption + lyrics & EMDC ablation \\
          M3a & 1.5B & on  & random local 30 s crop  & caption + lyrics & full-context ablation \\
          M3b & 1.5B & on  & full song & learned \texttt{null\_text} & decoder-side
          text ablation \\
          \bottomrule
      \end{tabular*}
      \endgroup
  \end{table}

  \begin{table}[t]
      \centering
      \caption{Controlled FullDiT ablation results. Upward and downward arrows
      indicate whether higher or lower values are better. Boldface marks the
      best objective result in each row. Song-level preference reports the win
      rate of each ablated model against M1 over non-tied songs.}
      \label{tab:fulldit_ablation_results}
      \begingroup
      \scriptsize
      \setlength{\tabcolsep}{2.5pt}
      \renewcommand{\arraystretch}{1.08}
      \begin{tabular*}{\linewidth}{@{\extracolsep{\fill}}llcccc@{}}
          \toprule
          \textbf{Evaluation regime} &
          \textbf{Metric} &
          \textbf{M1} &
          \textbf{M2} &
          \textbf{M3a} &
          \textbf{M3b} \\
          \midrule
          clean GT-codec
              & ViSQOL $\uparrow$ & 3.3388 & \textbf{3.4578} & 2.8863 & 3.3021 \\
          & log-mel L1 $\downarrow$ & 1.1363 & \textbf{0.9248} & 1.5381 & 1.1865 \\
          & MR-STFT distance $\downarrow$ & \textbf{1.7806} & 1.7896 & 1.8707 & 1.8253 \\
          \midrule
          synthetic-corrupted GT-codec
              & ViSQOL $\uparrow$ & \textbf{3.2036} & 2.4342 & 2.6868 & 3.1838 \\
          & log-mel L1 $\downarrow$ & \textbf{1.1842} & 2.2372 & 2.0034 & 1.2069 \\
          & MR-STFT distance $\downarrow$ & \textbf{1.7943} & 2.3346 & 2.0190 & 1.8327 \\
          \midrule
          LM-generated codec
              & Audiobox PQ $\uparrow$ & \textbf{8.213} & 8.122 & 6.011 & 8.135 \\
          & song-level preference vs. M1 $\uparrow$
              & Ref. & 30.3\% & 0.0\% & 0.0\% \\
          \bottomrule
      \end{tabular*}
      \endgroup
  \end{table}

\paragraph{Evaluation regimes.}
  We evaluate the four models under three complementary codec conditions. For
  \emph{clean GT-codec}, a frozen tokenizer extracts ground-truth codec tokens
  from each held-out song, which FullDiT uses to reconstruct the paired source
  audio. For \emph{synthetic-corrupted GT-codec}, we apply the same fixed
  EMDC-targeted corruption to these ground-truth tokens, using codebook-wise
  replacement rates and cosine-KNN near-miss replacements while retaining the
  original song as the acoustic target. This setting isolates robustness to
  controlled codec-interface errors. For \emph{LM-generated codec}, the frozen
  upstream hybrid-LM predicts codec sequences from the caption and lyrics. Since
  such sequences admit multiple valid acoustic realizations, this regime has no
  unique reference waveform. We fix the prompts, lyrics, LM-generated tokens,
  song lengths, random seeds, and sampling recipe across M1--M3b.

  \paragraph{Metrics and sample sizes.}
  All objective results use $N=200$ samples: clean and synthetic-corrupted
  GT-codec share the same 200 held-out songs, while LM-generated codec uses a
  fixed set of 200 generations. For the two GT-codec regimes,
  ViSQOL~\cite{chinen2020visqol} measures full-reference perceptual similarity,
  with higher values indicating closer perceptual agreement with the source.
  Log-mel L1 is the mean absolute difference between the generated and reference
  log-mel spectrograms and measures spectral-envelope mismatch, while
  multi-resolution STFT distance~\cite{yamamoto2020parallelwavegan} compares
  magnitude spectra at multiple time--frequency resolutions; lower values are
  better for both distances. For LM-generated codec, we use Audiobox Aesthetics
  Production Quality (PQ)~\cite{tjandra2025audiobox}, a learned no-reference
  estimate of technical production quality, for which higher is better. We
  additionally conduct blinded pairwise listening tests on a fixed set of
  $N=100$ songs with five expert raters with musical backgrounds. M1 serves as
  the common reference in comparisons with M2, M3a, and M3b, and the reported
  song-level preference is
  $\mathrm{wins}/(\mathrm{wins}+\mathrm{losses})$ after excluding ties. All four
  models use the same inference recipe, including the selected four-way CFG
  setting
  $(s_{\mathrm{codec}},s_{\mathrm{lyrics}},s_{\mathrm{caption}})=(1,2,1)$.

  Table~\ref{tab:fulldit_ablation_results} reveals three consistent trends.
  First, removing EMDC improves two clean-codec reconstruction metrics but
  substantially degrades all metrics under synthetic corruption; M1 also obtains
  higher PQ and is preferred to M2 under LM-generated codec conditions. This
  trade-off indicates that EMDC improves robustness to imperfect codec inputs
  rather than merely optimizing clean reconstruction. Second, M3a, which
  replaces full-song training with random 30-second crops, incurs the largest PQ
  drop and is consistently worse than M1, highlighting the importance of
  full-song context for acoustic rendering. Finally, removing decoder-side
  caption and lyrics produces smaller objective degradations, but M3b remains
  consistently below M1 and receives no wins in the non-tied song-level
  comparison, showing that text provides useful information beyond the codec
  sequence.

  \FloatBarrier

\subsection{Fine-grained Automatic Evaluation}

To evaluate the effectiveness of the proposed system, we report complementary automatic evaluations using SongBench~\cite{wu2026songbench}, SongEval~\cite{yao2025songeval}, AudioBox-Aesthetic~\cite{tjandra2025audiobox}, and CMI-Reward~\cite{ma2026cmirewardbench}. The independent preference-based evaluation is presented in the Introduction.

\begin{table}[!htbp]
    \centering
    \caption{Automatic evaluation results for multilingual vocal music generation on a test set containing 500 examples, balanced across eight major genres and five languages (Chinese, English, Japanese, Korean, and Spanish). We report mean scores from SongBench, SongEval, AudioBox-Aesthetic, and CMI-Reward. Higher scores indicate better evaluator outputs except for AudioBox Production Complexity, which is a descriptive measure and is not inherently better when larger. The highest point estimate in each row is shown in bold.}
    \label{tab:fine_grained_automatic_evaluation}
    \begingroup
    \scriptsize
    \setlength{\tabcolsep}{1.5pt}
    \renewcommand{\arraystretch}{1.00}
    \begin{tabular*}{0.90\linewidth}{@{\extracolsep{\fill}}lcccccc@{}}
        \toprule
        \textbf{Metrics} &
        \textbf{Ours} &
        \textbf{Suno V5.5} &
        \textbf{Suno V5} &
        \textbf{Mureka V8} &
        \textbf{Lyria 3 Pro} &
        \textbf{MiniMax Music 2.6} \\
        \midrule
        \multicolumn{7}{c}{\emph{SongBench}} \\
        \midrule
        Melody       & \textbf{7.2001} & 6.6939 & 6.9202 & 7.0660 & 7.0377 & 6.7245 \\
        Arrangement  & \textbf{7.3879} & 6.8406 & 7.0805 & 7.2601 & 7.1511 & 6.8167 \\
        Musicality   & \textbf{6.3678} & 5.8297 & 6.0096 & 6.2559 & 6.1810 & 5.8823 \\
        Vocal        & 7.6234 & 7.0981 & 7.3160 & \textbf{7.6248} & 7.4560 & 7.2513 \\
        Instrumental & \textbf{7.3517} & 7.0080 & 7.1487 & 7.2585 & 7.0583 & 6.8458 \\
        Mixing       & \textbf{7.3047} & 7.0332 & 7.1101 & 7.1433 & 7.0093 & 6.7397 \\
        Structure    & 6.9650 & 6.6158 & 6.7363 & 7.0237 & \textbf{7.0565} & 6.5637 \\
        \midrule
        \multicolumn{7}{c}{\emph{SongEval}} \\
        \midrule
        Coherence    & \textbf{4.5094} & 4.2905 & 4.3278 & 4.3870 & 4.4673 & 4.2772 \\
        Musicality   & \textbf{4.4098} & 4.1547 & 4.1970 & 4.2875 & 4.3466 & 4.1781 \\
        Memorability & \textbf{4.4663} & 4.2123 & 4.2473 & 4.3680 & 4.4153 & 4.2180 \\
        Clarity      & \textbf{4.3838} & 4.1246 & 4.1741 & 4.2431 & 4.3099 & 4.1448 \\
        Naturalness  & \textbf{4.2600} & 4.0046 & 4.0679 & 4.1767 & 4.2021 & 3.9919 \\
        \midrule
        \multicolumn{7}{c}{\emph{AudioBox-Aesthetic}} \\
        \midrule
        Content Enjoyment     & \textbf{7.7089} & 7.5289 & 7.5507 & 7.5921 & 7.6950 & 7.6844 \\
        Content Usefulness    & \textbf{7.9989} & 7.9535 & 7.8020 & 7.8634 & 7.8919 & 7.9578 \\
        Production Complexity & \textbf{6.8774} & 6.6486 & 6.8118 & 6.8768 & 6.7561 & 6.7514 \\
        Production Quality    & \textbf{8.2986} & 8.2083 & 8.1411 & 8.1098 & 8.2780 & 8.2923 \\
        \midrule
        \multicolumn{7}{c}{\emph{CMI-Reward}} \\
        \midrule
        Alignment & 2.1786 & 1.9216 & 1.9119 & \textbf{2.3089} & 2.0377 & 2.1546 \\
        Quality   & \textbf{2.7611} & 2.3665 & 2.3756 & 2.7590 & 2.5280 & 2.6661 \\
        \bottomrule
    \end{tabular*}
    \endgroup
\end{table}

Table~\ref{tab:fine_grained_automatic_evaluation} compares our system with five
representative commercial music generation systems on the same 500 test cases,
balanced across eight major genres and five languages: Chinese, English,
Japanese, Korean, and Spanish. Across the four evaluator families, our system
achieves the highest mean score in 15 of 18 reported dimensions: five of seven
SongBench dimensions, all five SongEval dimensions, all four
AudioBox-Aesthetic dimensions, and CMI-Reward Quality.

On SongBench, our system leads in Melody, Arrangement, Musicality, Instrumental,
and Mixing. The corresponding point estimates suggest richer and more memorable
melodic lines, stronger harmonic and instrumental orchestration, greater
overall artistic appeal, more realistic instrument rendering, and better track
balance and spatial imaging. Its Vocal score (7.6234) nearly matches the best
result from Mureka V8 (7.6248), while the remaining gap on Structure highlights
room for improving section organization and transitions.

Our system obtains the highest mean score in every SongEval dimension.
Coherence and Clarity reflect continuity across song sections and clearly
organized musical form; Memorability captures distinctive melodies, rhythmic
motifs, and lyrical hooks; Naturalness specifically evaluates vocal breathing
and phrasing; and Musicality summarizes melody, harmony, instrumentation, and
vocal--accompaniment integration. This consistent profile accords with the
intended long-range modeling objective of hybrid-LM and
the perceptual goals of preference alignment. On AudioBox-Aesthetic, our system also
leads in Content Enjoyment, Content Usefulness, and Production Quality, while
exhibiting the highest Production Complexity. These outputs respectively
indicate predicted listening appeal, utility as creative source material,
technical production quality, and richer component layering; Production
Complexity is descriptive rather than a direct measure of quality. Finally,
our system attains the highest CMI-Reward Quality score, whereas Alignment
remains below Mureka V8. Overall, the results suggest a balanced advantage in
composition, vocal and instrumental realization, mixing, and final rendering,
while identifying condition following and explicit structural planning as
directions for further improvement.

\section{Conclusion}

In this report, we introduced a unified framework for controllable full-song generation. The proposed system supports Lyrics-to-Song Generation, Instrumental Music Generation, and Cover Song Generation within a single framework. To address the complexity of music representation, it adopts a semantic-aware 8-codebook RVQ tokenizer and hybrid-LM for discrete token generation. For melody-guided cover generation, the system further incorporates a two-level melody representation to capture both the main melodic contour and fine-grained pitch variation. To achieve high-fidelity synthesis for complete songs, FullDiT performs full-song flow matching with non-causal self-attention over a continuous VAE latent, using the full-length codec sequence, text caption, and lyrics when available as conditions. We further investigate DPO, GRPO, and OPD for reward-based hybrid-LM post-training and flow-based GRPO for FullDiT refinement. The proposed system demonstrates competitive performance on the multilingual automatic benchmark and the external Artificial Analysis Music with Vocals leaderboard evaluated in this report. Overall, our framework bridges discrete music composition, melody control, and full-length high-fidelity synthesis, offering a practical solution for high-quality song generation.

\FloatBarrier



\section*{Acknowledgements}

We thank Hongzhi Cai, Jiayan Cui, Gang Qiao, Zhicheng Tian,
Haixin Wang, Cheng Wen, Bajian Xiang, Jixing Yu, Peiyun Zeng, Sitong Zhao,
Tianyu Zhao, Qixi Zheng and Xin Zhou for their valuable contributions to data curation, system development, evaluation, and infrastructure support.
Names are listed alphabetically by family name.

\bibliographystyle{plain} 
\bibliography{reference}

\end{document}